\documentclass[twocolumn, prl, amssymb, superscriptaddress, aps, showpacs,preprintnumbers,
amsmath,showkeys,floatfix]{revtex4}

\usepackage{epstopdf}
\usepackage{graphics}
\usepackage{graphicx}
\usepackage{dcolumn}
\usepackage{bm}
\usepackage{longtable}
\usepackage{epsfig}
\usepackage{times}
\usepackage{url}
\usepackage{color}
\usepackage{subfigure}

\begin{document}

\title{Stopping narrow-band x-ray pulses in nuclear media}

\author{Xiangjin \surname{Kong}}
\email{xjkong@mpi-hd.mpg.de}
\affiliation{Max-Planck-Institut f\"ur Kernphysik, Saupfercheckweg 1, D-69117 Heidelberg, Germany}
\affiliation{Department of Physics, National University of Defense Technology, Changsha 410073, People's Republic of China}

\author{Adriana \surname{P\'alffy}}
\email{palffy@mpi-hd.mpg.de}
\affiliation{Max-Planck-Institut f\"ur Kernphysik, Saupfercheckweg 1, D-69117 Heidelberg, Germany}

\date{\today}

\begin{abstract}
A control mechanism for stopping x-ray pulses in resonant nuclear media is investigated theoretically. We show that narrow-band x-ray pulses can be mapped and stored as nuclear coherence in a thin-film planar x-ray cavity with an embedded $^{57}\mathrm{Fe}$  nuclear layer. The pulse is nearly resonant to the 14.4 keV M\"ossbauer transition in the $^{57}\mathrm{Fe}$ nuclei. The role of the control field  is played here by a hyperfine magnetic field which induces interference effects reminding of electromagnetically induced transparency. We show that by switching off the control magnetic field, a narrow-band x-ray pulse can be completely stored in the cavity for approximately hundred ns.  Additional manipulation of  the external magnetic field can lead to both group velocity and phase control of the pulse in the x-ray cavity sample.

\end{abstract}

\maketitle

Recent years have witnessed the  commissioning of coherent x-ray sources opening the new field of x-ray quantum optics \cite{adams2013}. While not yet as advanced as its optical counterpart, the latter may enable coherent control of x-rays, with potential applications for the fields of metrology, material science, quantum information, biology and chemistry. The desirable properties of x-rays are deeper penetration, better focus, no longer limited by an inconvenient diffraction limit as for optical photons, correspondingly spatial resolution, robustness,  and the large momentum transfer they may produce. A peculiar circumstance is that x-rays are resonant to either inner shell electron transitions in (highly) charged ions \cite{Young2010.N,Kanter2011.PRL,Rohringer2012}, or transitions in atomic nuclei \cite{Buervenich2006,rohlsberger2012electromagnetically}. First experiments towards the demonstration   of nonlinear phenomena with x-rays have been performed with atoms \cite{XPDC1,XPDC2,FuchsPRL,FuchsNPhys} and nuclei \cite{rohlsberger2010collective,rohlsberger2012electromagnetically,heeg2013vacuum,Fano2015,slowlight2015,Vagizov2014,Olga2015}. 
Nonlinear interactions between x-rays and nuclei are a promising candidate to control x-ray pulses, which remains challenging so far \cite{slowlight2015,Fano2015,rohlsberger2012electromagnetically,PhysRevLett.77.3232}.

High-performance control over x-rays is compulsory if also x-ray qubit applications in quantum information or cryptography are to be realized \cite{Olga2015,JonasArXiv2015}, such as, for instance, preparation of entangled ensembles \cite{PhysRevLett.84.4232}, generation of squeezed states \cite{PhysRevLett.83.1319}, quantum memories \cite{julsgaard2004experimental,Hosseini2009} or photonic circuits \cite{almeida2004all,liu2010,shadbolt2012generating,moss2013}, already accomplished in the long-wavelength regime. Main difficulties  compared to the optical regime are the lack of high quality factor cavities and of suitable level schemes that would facilitate established control schemes.

In this Letter, we demonstrate from the theory side that a spectrally narrow x-ray pulse can be mapped and stored as nuclear coherence in a thin-film planar x-ray cavity \cite{Roehlsberger2004} with embedded layers containing nuclei with a transition resonant to the x-ray pulse. This novel storage mechanism relies on interference effects possible due to the occurrence of spontaneously generated coherences specific to the nuclear system \cite{heeg2013vacuum}. We lay out the theoretical formalism for describing this system and show that storage can be described by the formation of a dark-state polariton \cite{fleischhauer2000dark} and the coherent control over the polariton matter and radiation parts is provided by an external magnetic field. Surprisingly, despite the very different level scheme and applied fields,
the dynamics of the x-ray cavity with an embedded nuclear layer in the presence of a hyperfine magnetic field is governed by equations reminiscent of electromagnetically induced transparency (EIT) in atomic media \cite{EITReview2005}. 
Our results show that a spectrally narrow x-ray pulse can be completely mapped onto nuclear coherences and retrieved at later times, with storage time determined by the nuclear excited state mean lifetime, on the order of hundred ns. Our scheme is based on a different operation principle than previously implemented or proposed storage setups \cite{PhysRevLett.77.3232,liao2012coherent}, with the two major advantages that (i) it is more reliable and much easier to implement experimentally  and (ii) it works for a broader spectrum of parameters, e.g., storage times or variable pulse width. We anticipate this setup can become a versatile tool for control of spectrally narrow x-ray pulses.

EIT is a quantum interference effect that can be used to render a resonant opaque medium transparent. Typically, EIT can be achieved in a so-called $\Lambda$ three-level system  driven simultaneously by a strong control and a weaker probe pulse. Due to the control field, the medium becomes transparent for the probe pulse in a narrow window around the resonance frequency \cite{EITReview2005}.  In the optical regime, EIT can be used to slow down  \cite{lHau1999,Baba2008,Wu2010} and even to stop light in an atomic medium \cite{phillips2001storage,liu2001observation,LukinRev2003,Gorshkov2007,Halfmann2013} by a sudden turning off of the control field. 
However, due to the lack of two-color x-ray sources \cite{twocolSACLA,twocolLCLS} and the proper nuclear three-level systems, the traditional optical EIT scenario is not available at present for x-rays. So far, an alternative setup with two nuclear layers in a thin-film planar x-ray cavity has reported EIT-like behavior with nuclear transitions \cite{rohlsberger2012electromagnetically}. In addition, in a recent work \cite{slowlight2015} slow light was observed in the hard x-ray regime 
by introducing a steep linear dispersion in the nuclear optical response. 

\begin{figure}[t]
\centering
\includegraphics[width=0.5\textwidth]{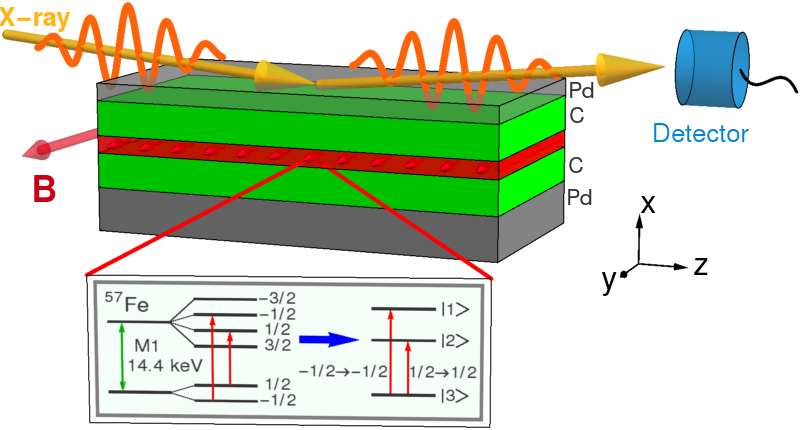}
\caption{\label{fig1} (color online). Thin-film planar cavity setup with x-ray grazing incidence. The cavity consists of a sandwich of Pd and C layers with a  1~nm layer containing $^{57}\mathrm{Fe}$  placed at the antinode of the cavity. The nuclei experience a hyperfine magnetic field $\vec{B}$  (red horizontal arrow). Inset panel: $^{57}\mathrm{Fe}$ level scheme with hyperfine splitting. This is equivalent with a $V$-like three-level scheme comprising the common ground state  and the two excited states (\ref{excited}).}
\end{figure}

In this Letter, we investigate an EIT-like behavior 
based on the phenomenon of spontaneously generated coherence \cite{heeg2013vacuum}.  Instead of a strong control field to produce a splitting of the excited state, we envisage a hyperfine magnetic field perpendicular to the x-ray propagation direction $\hat{k}$  that induces the hyperfine splitting of the ground and excited $^{57}\mathrm{Fe}$ nuclear states \cite{heeg2013vacuum}. The stable ground state of $^{57}$Fe (nuclear spin $I_g=1/2$) is then split into a doublet with $m_g=\pm1/2$ and the first excited state at 14.4 keV (nuclear spin $I_e=3/2$, mean lifetime $\tau_0$=141~ns) into a quadruplet with $m_e=\pm1/2,\pm3/2$. A single $^{57}\mathrm{Fe}$ layer placed in a planar cavity for hard x-rays similar to the setup in Ref.~\cite{rohlsberger2010collective} is probed by the x-ray pulse at grazing incidence, as shown in Fig.~\ref{fig1}. Depending on the polarization of the incident light, specific transitions between the six hyperfine-split nuclear states will be driven. In the following, we consider a linearly polarized x-ray pulse such that only the two $\Delta m=m_e-m_g=0$ transitions can be driven. The nuclear scattering response is measured in the reflected signal at the detector.

We describe the hyperfine level scheme in terms of the collective states of the $N$-nuclei ensemble in the cavity. Initially, the nuclei are in the collective ground state $|G\rangle$ with $N_i$ nuclei in the $|g_{i}\rangle$ ground state sublevel, where $i\in{1,2}$, $N_1+N_2=N$ and $N_1\approx N_2$ at room temperature. Typically, in synchrotron radiation (SR) pulses at most only one photon will be resonant to the $^{57}\mathrm{Fe}$ nuclear transition.
 Due to the recoilless nature of the M\"ossbauer nuclear transition in solid-state nuclear targets, a delocalized, 
collective excitation (in literature referred to as  nuclear exciton \cite{NFS-bible} or  timed Dicke state \cite{Scully2009.superradiance}) will be created by the single resonant photon. 
We define the excited state as a timed Dicke state  \cite{Scully2009.superradiance}
\begin{equation}
 | E_{\mu} \rangle =\frac{1}{\sqrt{N_{\mu}}}\sum_{n}^{N_{\mu}}e^{i\vec{k}_C \cdot \vec{R}^{(n)}}|g_{1}^{(1)}\rangle \ldots |e_{\mu}^{(n)} \rangle \ldots | g_{2}^{(N)}\rangle
\label{excited}
\end{equation}
in which the $n$th atom has been excited by the transition $\mu$, with the notation $\mu=1$ for the transition $m_g=-1/2\rightarrow m_e=-1/2$ and $\mu=2$ for $m_g=1/2\rightarrow m_e=1/2$, depending on the initial ground state spin projection $m_g$. The position of the excited nucleus is given by $\vec{R}^{(n)}$ and $\vec{k}_C$ represents the total wave vector for the resonant cavity mode. The two $\Delta m=0$ transitions are equivalent in this system to 
the two transitions $|3\rangle \rightarrow |1\rangle$ and $|3\rangle \rightarrow |2\rangle$, where we have used the notation $|3\rangle$ for  $ | G \rangle$,  $|1\rangle$ for $| E_{1} \rangle $  and $|2\rangle$ for $| E_{2} \rangle$, respectively, as illustrated in the inset panel in Fig.~\ref{fig1}. The two transitions experience  vacuum-mediated coupling by spontaneously generated coherence terms  \cite{heeg2013vacuum,slowlight2015}.

The theoretical formalism used to investigate the  system dynamics  extends the quantum optical model for x-ray thin film cavities in Ref.~\cite{heeg2013x}. The dynamics of the system is described by the master equation \cite{ScullyZubairy}
$ \frac{d}{dt}\rho=-i[\hat{H},\rho]+\mathcal{L}[\rho]$, which has proved to be very successful in modelling the interaction of SR with nuclei in bulk samples or thin-film cavities \cite{Shvydko1999,Kong2014,heeg2013x}. Here, $\hat{H}$  is the total system Hamiltonian and the Lindblad operator $\mathcal{L}$ accounts for the cavity loss  rate $\kappa$  and the nuclear spontaneous emission rate $\gamma$ which is connected to the nuclear mean lifetime $\tau_0$ by $\gamma=\hbar/\tau_0$. With $\hbar=1$, in  the following $\gamma$ is used as both rate and width. The observable in the system is the reflection coefficient which is obtained in the bad-cavity limit, i.e., 
the decay rate of the cavity $\kappa$ is much larger than the atom-field cavity coupling strength $g$. Under the steady-state conditions  \cite{heeg2013x}, the 
reflection coefficient is proportional to the coherence terms $\langle E_\mu|\rho|G\rangle$ (see also  Supplementary Material \cite{supplmat} for detailed analytical derivations). 
The sum of the two relevant coherences in the system is given by
\begin{equation}
 \rho_{23}+\rho_{13}=\frac{i\sqrt{\frac{16}{3}}g\sqrt{N}\Omega(\gamma-2i\Delta)}{(\gamma-2i\Delta)(\gamma'-2i\Delta')+(2\phi)^2}\, ,
\label{cohsum}
\end{equation}
where $\Delta$ is the detuning between the x-ray field and the single-nucleus transition energy of the nuclei in the absence of hyperfine splitting. The primed quantities $\gamma'=\gamma+\frac{4}{3}\lvert g \rvert^2N\zeta_S$, $\Delta'=\Delta-\frac{2}{3}\lvert g \rvert^2N\delta_{LS}$ stand for the collective, cavity-enhanced decay rate and detuning, respectively with $\delta_{LS}=-\frac{\Delta_c}{\kappa^2+\Delta_c^2}$ and $\zeta_S=\frac{\kappa}{\kappa^2+\Delta_c^2}$, where $\Delta_c$ is the cavity detuning proportional to the x-ray incident angular deviation from the resonant angle $\varphi_0$, see also  Supplementary Material \cite{supplmat}. Furthermore, in Eq.~(\ref{cohsum}) the parameter $\Omega$ is given by $\Omega=\frac{\sqrt{2\kappa_R} a_{in}}{\kappa+i\Delta_c}$, where  $\kappa_{R}$ denotes the x-ray coupling strength into the cavity mode, and $a_{in}$ characterizes the driving field of the cavity mode by the external (classical) x-ray field. The last term in the denominator in Eq.~(\ref{cohsum}) is determined by $\phi=\frac{1}{2}(\delta_g+\delta_e)$, where $\delta_g$ ($\delta_e$) denotes the hyperfine energy difference between two adjacent ground (excited) sub-states. 

\begin{figure}[t]
\centering
\includegraphics[width=0.40\textwidth]{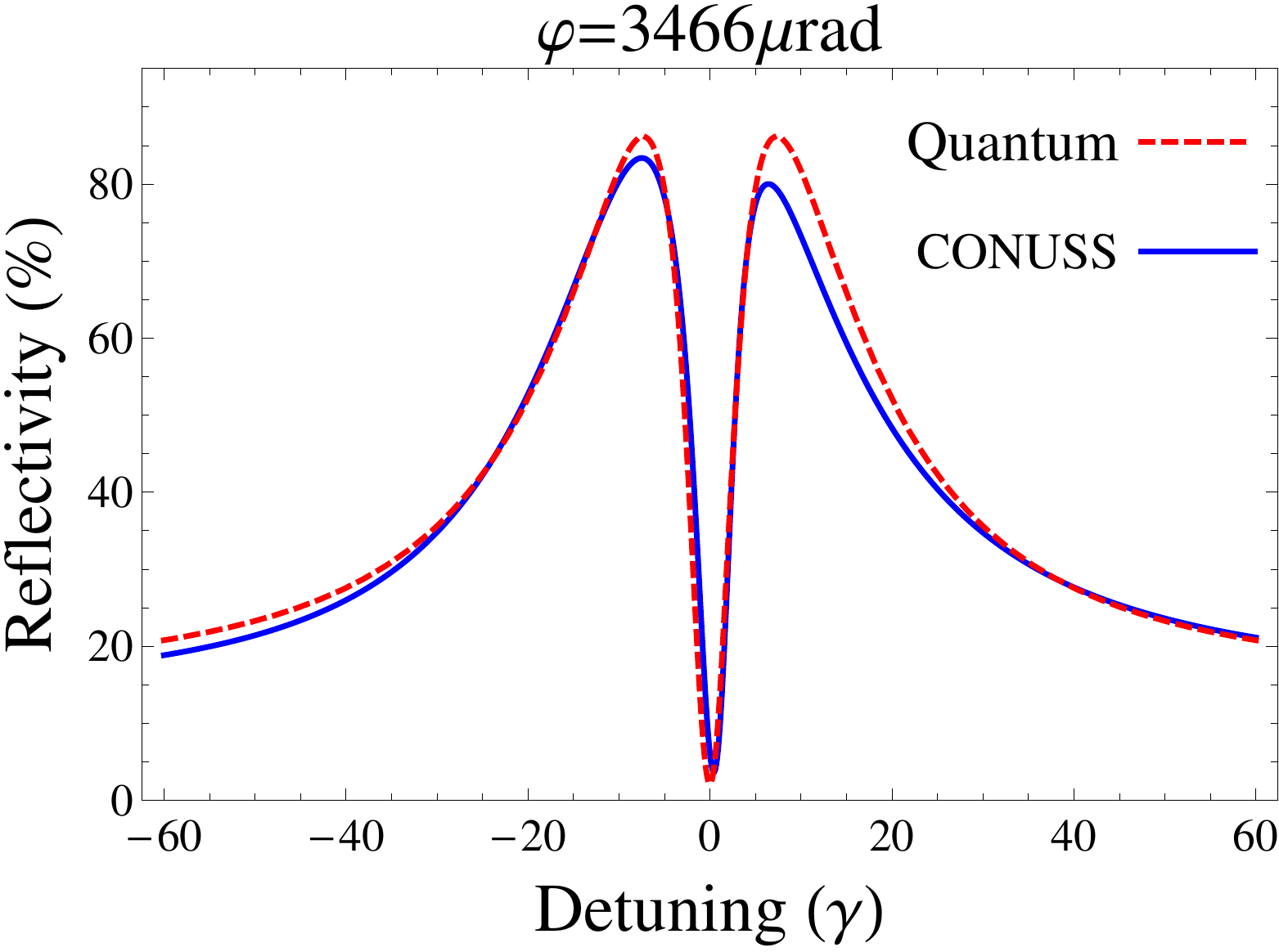}
\caption{\label{fig2} (color online). Reflectivity spectra calculated with CONUSS (solid blue line) and the quantum cavity model (dashed red line) for the incident angle $\varphi$. The considered cavity structure is Pd(5nm)/C(20nm)/Fe(1nm)/C(20nm)/Pd(30nm) and $B=$6.4 T.  }
\end{figure}

Numerical results for the reflectivity of the cavity presented in Fig.~\ref{fig1} in an external hyperfine magnetic field with $B=$6.4~T for the 
 resonant case ($\Delta_c=0$, corresponding to  $\varphi=\varphi_0=3466$ $\mu$rad) as a function of the x-ray detuning  are presented in Fig.~\ref{fig2}. The hyperfine magnetic field introduces the energy splitting $\phi=6\gamma$. Further parameter values are $\kappa_R=3.1\cdot 10^5\gamma,\,\kappa=4.6\cdot 10^5\gamma,\,\sqrt{N}\lvert g \rvert=2500\gamma$. A dip  very similar to the well-known EIT absorption spectra in the atomic case and to the 
x-ray EIT results presented in Ref.~\cite{rohlsberger2012electromagnetically} can be observed. We note that for the resonant case, $\delta_{LS}=-\frac{\Delta_c}{\kappa^2+\Delta_c^2}=0$, $\Delta'=\Delta$ and with $\gamma'\gg \gamma$,
 the expression of  the coherence sum $\rho_{21}+\rho_{31}$ in Eq.~(\ref{cohsum}) is very similar to the EIT coherence \cite{EITReview2005}, see also Supplementary Material \cite{supplmat}. Thus, the reflectivity of the thin-film cavity behaves analogously to the EIT absorption and the transmission is the equivalent of atomic medium transparency. We use the software package CONUSS \cite{sturhahn2000conuss} implementing a self-consistent theory including multiple scattering to all orders \cite{Ralf1999} to verify our reflection spectra. 
 Theoretical predictions by CONUSS have proven to agree extremely well with  experimental results \cite{rohlsberger2010collective,rohlsberger2012electromagnetically,heeg2013vacuum,Fano2015} such that they are often used  as benchmark. The comparison  in Fig.~\ref{fig2} shows that the quantum formalism used is very reliable in describing the transparency window.

Next we investigate how to store an x-ray pulse exploiting the EIT-like features of the setup. The expression in Eq.~(\ref{cohsum}) reveals that the energy difference between the two transitions $2\phi$ plays the role of  control field from the traditional EIT case. In analogy with the atomic case where 
switching off the control field leads to pulse storage, we investigate the case when the hyperfine magnetic field in our cavity system disappears. 
In the slowly varying amplitude approximation for the x-ray field Rabi frequency $\Omega(z,t)$, and neglecting in this step the spontaneous decay of the system $\gamma$, we obtain for the propagation equation in the $z$-direction in the perturbative and adiabatic limit, see also Supplementary Material \cite{supplmat},
\begin{eqnarray}
 \left(\frac{\partial}{\partial t}+c\frac{\partial}{\partial z}\right)\Omega(z,t)&=& ig\sqrt{\frac{1}{3}N}[\rho_{32}(z,t)+\rho_{31}(z,t)] \nonumber \\
&\approx& 
-\frac{2g^2N}{3\phi(t)}\frac{\partial }{\partial t}\frac{\Omega(z,t)}{\phi(t)}\, .
\label{coherences}
\end{eqnarray}
The group velocity of the x-ray pulse is smaller than the light velocity  in vacuum $c$ according to $v_g=c/(1+\frac{2g^2N}{3\phi^2})$. Such subluminal x-ray propagation in thin-film cavities has been confirmed experimentally in Ref.~\cite{slowlight2015}. We may introduce here the dark-state polariton originally studied in the atomic case \cite{fleischhauer2000dark}
\begin{eqnarray}
\Psi(z,t)&=&{\rm cos}\, \theta(t)\Omega(z,t) \nonumber \\
&-& {\rm sin}\,\theta(t)\sqrt{\frac{2N}{3}}\left[\rho_{31}(z,t)-\rho_{32}(z,t)\right]  \, ,
\label{polariton}
\end{eqnarray}
with ${\rm cos}\,\theta(t)=\frac{\phi(t)}{\sqrt{\phi^2(t)+\frac{2}{3}g^2N}}$ and ${\rm sin}\,\theta(t)=\frac{g\sqrt{\frac{2}{3}N}}{\sqrt{\phi^2(t)+\frac{2}{3}g^2N}} $.  The polariton dynamics is governed by the expression
\begin{equation}
\left[\frac{\partial}{\partial t}+c\, {\rm cos}^2\,\theta(t)\frac{\partial}{\partial z}\right]\Psi(z,t)=0 \, ,
\label{poldyn}
\end{equation}
which describes a shape-preserving propagation with velocity $v=v_g=c {\rm cos}^2\,\theta(t)$ \cite{fleischhauer2000dark},
\begin{equation}
 \Psi(z,t)=\Psi\left(z-c\int_{0}^t   {\rm cos}^2\,\theta(\tau) d\tau, t=0\right) \, .
\end{equation}
The expression above defines a shape-preserving, polariton-like mixture of  electromagnetic field and collective nuclear coherences. If the hyperfine magnetic field is switched off while the x-ray pulse is inside the medium, the propagation velocity of the polariton $v=c\frac{\phi^2(t)}{\phi^2(t)+\frac{2}{3}g^2N}$ reduces to zero. By switching on the magnetic field after some time, the polariton will resume propagation through the sample at the original velocity. The x-ray pulse has been transferred to the nuclear coherences and then back by the external operation on the hyperfine energy splitting in the system.

\begin{figure}[t]
\centering
\includegraphics[width=0.5\textwidth]{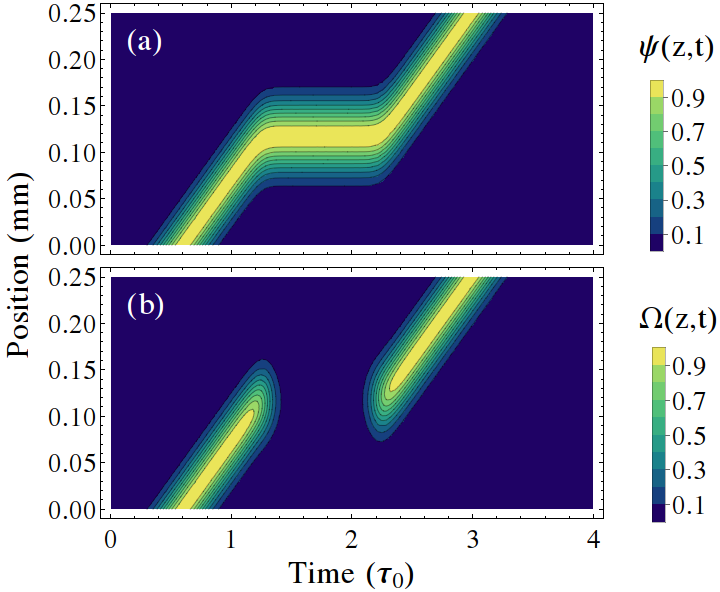}

\caption{\label{fig3} (color online). Propagation dynamics of the polariton field $\Psi$ (a) and the Rabi frequency $\Omega$ of the pulse (b) . The magnetic field is switched off at $ t=1.3\tau_0\,$ and switched back on at $ t=2.2\tau_0\,$.}
\end{figure}

Numerical results obtained from the evaluation of Eqs.~(\ref{coherences}-\ref{poldyn}) are presented in Figs.~\ref{fig3}. As incoming pulse we choose for illustration purposes a Gaussian-shape pulse  $\Omega_{p}(t,0)=\Omega_{p0}\exp[-(\frac{t}{t_0})^{2}]$ with a  $t_0=0.2\tau_0$, i.e., a bandwidth of $2\gamma$. The pulse enters the medium and undergoes spatial compression as the  velocity is diminishing from $c$ to $v_g$ via EIT. 
 In order to store the x-ray pulse, we switch off the magnetic field at $ t=1.3\tau_0\,$ after the entire pulse has entered the medium. As a result, the velocity of the quantum field $\Psi$ reduces to zero and the x-ray pulse is mapped onto the collective nuclear coherences. The process can be reversed by switching back on the hyperfine magnetic field, for instance at $ t=2.2\tau_0\,$. The polariton then resumes its propagation through the medium at the same velocity as before, as shown in Fig.~\ref{fig3}(a). The evolution of the photon field part of the polariton is depicted in Fig.~\ref{fig3}(b) and shows that during the storage, the pulse is completely mapped onto the nuclear coherences. Note that the storage time is limited by the incoherent nuclear decay rate $\gamma$ (corresponding to $\tau_0=141$ ns) that characterizes the nuclear coherences, leading to an exponential drop of the pulse intensity, for illustration purposes not included in Figs.~\ref{fig3}.

In practice, spectrally narrow pulses can be produced by a single-line resonant spectral analyzer and mechanical or polarization-based removal of  the non-resonant component \cite{slowlight2015}. 
Alternatively, a SR M\"ossbauer source based on  a narrow-band,
pure nuclear reflection off a $^{57}$FeBO$_{3}$ crystal \cite{MoessbauerSR,mitsui2007synchrotron,potapkin201257fe,toellner1995polarizer,toellner2011synchrotron} could be employed.
 This source  provides $^{57}$Fe resonant radiation at 14.4 keV within a bandwidth of 15 neV which is tunable in energy over a range of  about $\pm0.6$ $\mu$eV \cite{potapkin201257fe}.
 The temporal profile of the SR M\"ossbauer pulse is given by the nuclear scattering in $^{57}$FeBO$_{3}$ and presents a modulation determined by the Bessel function of the first kind, $\left( \frac{\xi}{\sqrt{\xi\gamma t}}J_{1}\left[2\sqrt{\xi\gamma t}\right]\right)^{2} e^{-\gamma t}$, with $\xi$ the optical depth of the sample \cite{liao2012coherent}. Our numerical results show that the exact initial shape of the pulse does not introduce dispersion in the EIT-based storage.  In addition, similarly to the atomic case \cite{adiabaticity2001,adiabaticity2002}, we expect the  adiabaticity condition to be relaxed and  not restrict the storage experimentally.
Finally, our theoretical simulation considers a switching time of approx. 50 ns for the hyperfine magnetic field. Turning on and off  the external magnetic fields of few Tesla on the ns time scale can be achieved by using small single- or few-turn coils and a moderate pulse current of approx.~15 kA from low-inductive high-voltage ``snapper'' capacitors \cite{Miura2003}. Another mechanical solution, e.g., the lighthouse setup \cite{Roehlsberger2000} could be used to move the excited target out of and into a region with confined static $\vec{B}$ field, as discussed in Ref.~\cite{liao2012coherent}.

As in the case of atomic EIT, the transparency-based x-ray storage has a number of tunable features. The velocity of the pulse can be controlled by the magnitude of the hyperfine magnetic field $|\vec{B}|$. The orientation of $\vec{B}$ can decide upon the phase of the x-ray pulse. A rotation of the releasing magnetic field compared to the initial direction can lead to a phase modulation of the signal. The most relevant example is a releasing magnetic field orientated antiparallel to the initial one. This  is equivalent to a transformation ${\rm cos}\,\theta(t) \rightarrow -{\rm cos}\,\theta(t)$. Since the polariton in Eq.~(\ref{polariton}) is shape-preserving, the corresponding equivalent transformation of the electric field leads to $\Omega(z,t) \rightarrow -\Omega(z,t)$, which yields a phase modulation of $\pi$.

A comparison with other x-ray  setups  highlights the advantages of our scheme. In the thin-film x-ray cavities experiment in Ref.~\cite{rohlsberger2012electromagnetically}, the EIT-like dip in the reflectivity is created by the presence of a second $^{57}\mathrm{Fe}$ layer in the thin film cavity. Due to the  fixed cavity layer geometry, this setup has no tunable parameters and cannot be used to stop the x-ray pulse in the medium.
 In  Ref.~\cite{slowlight2015}, where  subluminal x-ray propagation in thin-film cavities was accomplished, the EIT transparency regime was avoided and the hyperfine magnetic field kept constant.  Coherent x-ray storage has been pioneeringly demonstrated in a nuclear forward scattering setup \cite{PhysRevLett.77.3232} almost two decades ago. That scheme works in the absorption regime: the x-ray photon excites a nucleus whose decay is suppressed by a rotation of the hyperfine magnetic field. A similar concept using the complete removal of the hyperfine magnetic field has been proposed in Ref.~\cite{liao2012coherent}. Both setups rely on the storage of the x-ray field in a nuclear excited state and the manipulation of the magnetic field at predetermined, fixed times $t_s$ which guarantee destructive interference and suppression of the nuclear decay. However, the latter may still occur prior to $t=t_s$ such that storage is neither deterministic nor really efficient. For example, while waiting for the earliest switching time in  Ref.~\cite{liao2012coherent}, the nuclear excited state has already decayed with 70\% probability. 

We conclude that our storage setup in the transparency window reminiscent of EIT is more flexible and more reliable than existing or proposed storage methods for x-rays. It relies on a different physical mechanism---the mapping of the x-ray pulse onto nuclear coherences, it is deterministic, and can be performed at variable times.  We anticipate that bringing  x-ray pulses to a halt will establish concepts used in atomic physics for nuclear physics with x-rays and render possible the processing and control of x-ray polarization qubits \cite{JonasArXiv2015} or time-bin qubits \cite{Vagizov2014,Olga2015} for 
quantum information applications in the x-ray regime.

The authors would like to thank  K. Heeg and J. Evers for fruitful discussions. XK gratefully acknowledges  financial support from the China Scholarship Council.

\bibliography{vrst-x}

\end{document}